\documentclass[conference]{IEEEtran}
\IEEEoverridecommandlockouts
\usepackage{cite}
\usepackage{amsmath,amssymb,amsfonts}
\usepackage{algorithmic}
\usepackage{graphicx}
\usepackage{textcomp}
\def\BibTeX{{\rm B\kern-.05em{\sc i\kern-.025em b}\kern-.08em
    T\kern-.1667em\lower.7ex\hbox{E}\kern-.125emX}}

\makeatletter
\def\ps@IEEEtitlepagestyle{
  \def\@oddfoot{\mycopyrightnotice}
  \def\@evenfoot{}
}
\def\mycopyrightnotice{
  {\footnotesize
  \begin{minipage}{\textwidth}
  \centering
  Copyright~\copyright~2018 IEEE.  Personal use of this material is permitted.  Permission from IEEE must be obtained for all other uses, in any current or future media, including reprinting/republishing this material for advertising or promotional purposes, creating new collective works, for resale or redistribution to servers or lists, or reuse of any copyrighted component of this work in other works. \\
  DOI: 10.1109/ICSRS.2018.8688854\\
  URL: https://ieeexplore.ieee.org/document/8688854
  \end{minipage}
  }
}

\begin{document}

\title{Combination of Component Fault Trees and Markov Chains to Analyze Complex, Software-controlled Systems}

\author{\IEEEauthorblockN{Marc Zeller}
\IEEEauthorblockA{\textit{Siemens AG, Corporate Technology} \\
81739 Munich, Germany \\
marc.zeller@siemens.com}
\and
\IEEEauthorblockN{Francesco Montrone}
\IEEEauthorblockA{\textit{Siemens AG, Corporate Technology} \\
81739 Munich, Germany \\
francesco.montrone@siemens.com}
}

\maketitle

\begin{abstract}
Fault Tree analysis is a widely used failure analysis methodology to assess a system in terms of safety or reliability in many industrial application domains. However, with Fault Tree methodology there is no possibility to express a temporal sequence of events or state-dependent behavior of software-controlled systems. In contrast to this, Markov Chains are a state-based analysis technique based on a stochastic model. But the use of Markov Chains for failure analysis of complex safety-critical systems is limited due to exponential explosion of the size of the model.
In this paper, we present a concept to integrate Markov Chains in Component Fault Tree models. Based on a component concept for Markov Chains, which enables the association of Markov Chains to system development elements such as components, complex or software-controlled systems can be analyzed w.r.t.~safety or reliability in a modular and compositional way.
We illustrate this approach using a case study from the automotive domain.
\end{abstract}


\section{Introduction}
\label{sec:introduction}
%
The importance of safety-critical systems in many application domains of embedded systems, such as aerospace, railway, health care, automotive and industrial automation is continuously growing. Thus, along with the growing system complexity, also the need for safety assessment as well as its effort is increasing drastically in order to guarantee the high quality demands in these application domains. 
The goal of the safety assessment process is to identify all failures that cause hazardous situations and to demonstrate that their probabilities are sufficiently low. In the application domains of safety-critical systems the safety assurance process is defined by the means of safety standards (e.g.~the IEC 61508 standard \cite{iec61508}).
\emph{Fault Tree Analysis (FTA)} \cite{vesely1981fault} is a common top-down deductive approach to identify failure modes, their causes, and effects with impact on the system safety.
With \emph{Component Fault Trees (CFTs)} \cite{Kaiser2003} there is a model- and component-based methodology for FTA, which supports a modular and compositional safety analysis strategy.

However, when using Fault Tree methodology there is no possibility to express a temporal sequence of events or state-dependent behavior. But this  is especially needed for modern, complex systems that are increasingly controlled by software components. In contrast to this, \emph{Markov Chains (MCs)} \cite{asmussen2008applied,IEC61165} are a state-based analysis technique and enable safety and reliability analysis of systems with temporal or state-based behavior (e.g.~fault tolerance capabilities). However, the use of MCs to analyze systems in industrial practice is limited due to exponential explosion of the size of the model when modeling large-scale systems.
In order to overcome these limitations and to combine the advantages of CFTs and MCs, an integrated approach using a component concept for MCs is required.

In this work, we present a concept to integrate Markov Chains into the Component Fault Tree methodology. This concept is based on a generalized component concept for MCs.

These so-called \emph{Component Markov Chain (CMC) elements} can be associated to system development elements such as components and reused along with the respective development artifact. Thus, providing a modular and compositional safety analysis strategy for MCs which is similar to CFTs. Thereby, so-called \emph{Generalized Hybrid Component Fault Trees (GHCFTs)} consisting of a combination of CFT and CMC elements can be created to analyze complex, software-controlled systems w.r.t.~safety and reliability.

The remainder of this paper is organized as follows: In Sec.~\ref{sec:relatedwork} we summarize related work. Afterwards, the concept of MCs and CFTs is outlined in Sec.~\ref{sec:background}. Sec.~\ref{sec:cmc} describes the generalized component concept for Markov Chains. Based on this concept, we introduce Generalized Hybrid Component Fault Trees in Sec.~\ref{sec:hyCFT} and show how to analyze the combination of CFT and CMC elements qualitatively and quantitatively. We illustrate our approach in Sec.~\ref{sec:casestudy} using an exemplary case study from the automotive domains. The paper is concluded in Sec.~\ref{sec:summary}.

\section{Related Work}
\label{sec:relatedwork}
Existing concepts for modular Markov Chains \cite{S0218539311004044,4273022,Boudali2007} solely focus on the modular design and composition of Markov models. The combination of Markov Chains and Fault Trees is not considered in these approaches.

Other approaches to represent temporal sequences of events or in Fault Trees, such as \cite{4224161,Bouissou2003149,Zixian20111591,Prohaska2015} focus on the integration of Markov Chains in Fault Trees as a substitute of the Fault Tree's Basic Events. Thereby, a Basic Event is represented by a MC. However, such an approach is limited to MCs with exactly one error state. Moreover, according to these existing solutions a MC can only be associated to specific components (e.g.~sensor components), which do not depend on other components.

\emph{Dynamic Fault Trees} \cite{159800} or \emph{AltaRica} \cite{6622976} use their own and specific syntax to describe a safety-critical system, which is more powerful in terms of expressiveness than classic Fault Trees. However, such approaches are rarely used in industry, since safety experts must first be trained on the new methodology and migration strategies for existing models are missing yet.

\emph{State/Event Fault Trees (SEFTs)} \cite{Kaiser2007} combine the concept of fault trees with a graphical representation of state/event semantics that is similar to state charts. SEFTs are partitioned into components using the component concept of CFTs. Each component is in exactly one state at an instant of time. State transitions are triggered by internal or external events.
Although, SEFTs extend fault tree analysis with capabilities to model temporal or state-based behavior, for quantitative analysis the SEFTs need to be transformed into \emph{Deterministic and Stochastic Petri Nets}. This results in a state explosion problem. Besides from the missing support of tools for modeling SEFTs there is no evidence that this approach can be applied to analyze complex systems in industrial practice.

In this work, we present a general component concept for Markov Chains which can be combined with Component Fault Tree methodology and which is not restricted by the above indicated limitations. It can be integrated into the CFT methodology seamlessly and analyzed qualitatively or quantitatively using well-established algorithms.

\section{Background}
\label{sec:background}

\subsection{Markov Chains}
\label{sec:mc}
Markov Chains (MC) \cite{asmussen2008applied,IEC61165} are a top-down analysis technique. A MC represents various system states and the relationships among them. 
MCs are often described by a sequence of directed graphs, where the edges of the graph (the so-called transitions) are labeled by the rates for going from one state to another state (the so-called transition rate). The transition rate from one state to another is either a failure rate $\lambda$ or a repair rate $\mu$. Each state of a MC is mutually exclusive because at any given time, the system can be in only one of the states. In this paper, we restrict our approach to the use of \emph{Continuous Time Markov Chain (CTMC)} with constant transition rates (failure rate or repair rate). Please note, that in the following we refer to CTMC as Markov Chain (MC).

Especially, in state-based or fault tolerant (software) systems the safety assessment process and evaluation of such system may be more appropriately achieved by the application of the Markov technique than using Fault Trees.

\subsection{Component Fault Trees}
\label{sec:cft}
A \emph{Component Fault Tree (CFT)} is a Boolean model associated to system development elements such as components \cite{Kaiser2003}. It has the same expressive power as classic fault trees that can for example be found in \cite{vesely1981fault}. As classic fault trees, also CFTs are used to model failure behavior of safety-critical systems. This failure behavior is used to document that a system is safe and it can also be used to identify drawbacks of the design of a system.

In CFTs, a separate \emph{CFT element} is related to a component. Failures that are visible at the outport of a component are models using \emph{Output Failure Modes (OFMs)} which are related to the specific outport. To model how specific failures propagate from an inport of a component to the outport, \emph{Input Failure Modes (IFMs)} are used. The internal failure behavior that also influences the output failure modes is modeled using the Boolean gates such as \emph{OR} and \emph{AND} as well as \emph{Basic Events}. 

In addition, it is possible to have more than one top event within a CFT element, such as an accident when a primary failure coincides with the failure of a countermeasure, but only a system unavailability when the same primary failure occurs while the countermeasure is working. 
Thus, the tree structure is extended towards a \emph{Directed Acyclic Graph (DAG)} structure in the CFT methodology. This avoids the artificial splitting of common cause errors into multiple \emph{repeated events}. Instead, it is possible for more than one path to start from the same basic event or sub-graph.

\begin{figure}[htb]
\includegraphics[width=\linewidth]{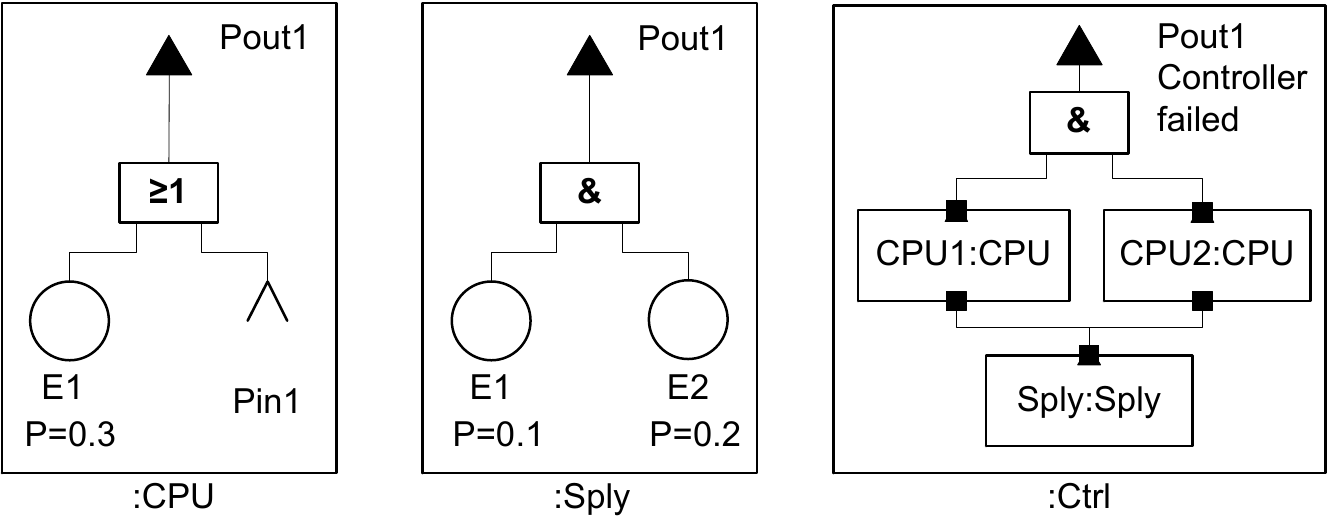}
\caption{Example of a simple CFT \cite{Kaiser2018}}
\label{fig:CFT_Introduction_Example}
\end{figure}

A small example of a CFT was presented in \cite{Kaiser2018} (see Fig.~\ref{fig:CFT_Introduction_Example}). 
The example shows an exemplary controller system \emph{Ctrl}, including two redundant \emph{CPU}s (i.e.~two instances of the same component type) and one common power supply \emph{Sply} (which would be a repeated event in traditional fault tree). The controller is unavailable if both CPUs are in the state "`failed"'. The inner fault tree of the component type CPU is shown on a separate screen; as the CPUs are of identical type, they only have to be modeled once and are then instantiated twice. 
The failure of a CPU can be caused by some inner basic event "`E1"' (the repetition of the ID "`E1"' in several components is not a problem, as each component constitutes its own name space). The failure of the CPU can also be caused by an external failure cause which is connected via an input port. As both causes result in a CPU failure, they are joined via a 2-input OR gate. 
The power supply is modeled as a separate component. Let us assume that the power supply is in its failed state if two separate basic failures are present (for example having two redundant batteries). 
Hence, instead of a single large fault tree, the CFT model consists of small, reusable and easy-to-review components.

Using this methodology of components also within fault tree models, benefits during the development can be observed in industrial practice, for example an increased maintainability of the safety analysis model \cite{KaiHofig2018}.

\section{A General Component Concept for Markov Chains}
\label{sec:cmc}
In this section, we present our general component concept for Continuous Time Markov Chains.
In the following, this method is described formally and illustrated using an example.

First, we assume that the System $S$ consists of a set of components $C = \left\{ c_{1},...,c_{n} \right\}$. Each component $c \in C$ includes a set of inports $IN(c) = \left\{ in_{1},...,in_{p} \right\}$ and a set of outports $OUT(c) = \left\{ out_{1},...,out_{q} \right\}$.
The information flow between the outport of a component $c_i \in C$ and the inport of another component $c_j \in C$ (with $c_i \neq c_j$) is represented by a set of connections
\begin{equation}
CON\left(c_i,c_j\right) = \left\{ (out, in) \; | \; out \in OUT(c_i) , in \in IN(c_j) \right\}
\end{equation}

A Markov Chain $MC$ is a directed graph which consists of a set of states $S = \left\{ s_i,...,s_n \right\}$, with an initial state $s_{init} \in S$ to start from and a set of error states $S_{error} \subset S$.
The relation between the states of the MC is defined by a set of transitions:
\begin{equation}
T = \left\{ (s_x, s_y) \; | \; s_x, s_y \in S \right\}
\end{equation}
Each transition $t(j,k) \in T$ from state $j \in S$ to state $k \in S$ is defined by a rate $\tau_{j,k}$, which either represents a constant failure rate $\lambda_{j,k}$ or a constant repair rate $\mu_{j,k}$:
\begin{equation}
t(j,k) := \tau_{j,k} = \left\{ 
\begin{array}{l l}
  \lambda_{j,k} \in \mathbb{R}^+ \\
  \mu_{j,k}     \in \mathbb{R}^+ \\ \end{array} \right.
\end{equation}

Hence, a Markov Chain is defined by the tuple
\begin{equation}
MC = \left( S, S_{error}, s_{init}, T \right)
\end{equation}

\begin{figure}[htbp]
  \centering
  \includegraphics[width=0.6\linewidth]{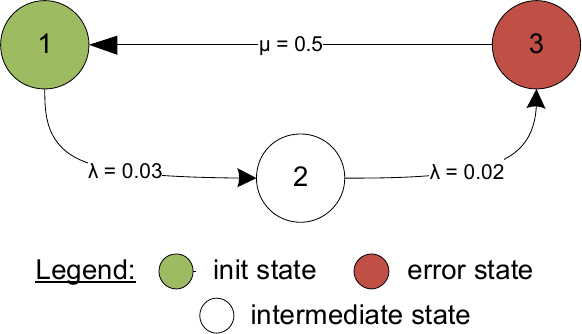}
  \caption{Exemplary Markov Chain (CMC)}
  \label{fig:example_MC}
\end{figure}

An example for a MC is illustrated in Fig.~\ref{fig:example_MC}.

This exemplary MC is defined by:
\begin{IEEEeqnarray*}{rCl}
S &=& \left\{ 1,2,3 \right\} \\
s_{init} &=& 1 \\
S_{error} &=& \left\{ 3 \right\} \\
T &=& \left\{ (1,2),(2,3),(3,1) \right\} \\
t(1,2) := \tau_{1,2} &=& \lambda_{1,2} = 0.03 \\
t(2,3) := \tau_{2,3} &=& \lambda_{2,3} = 0.02 \\
t(3,1) := \tau_{3,1} &=& \mu_{3,1} = 0.5 
\end{IEEEeqnarray*}

In order to specify a Component Markov Chain (CMC) $cmc_c$ which can be associated to any development artifact (e.g.~component) of the system $c \in C$, the definition of a MC must be extended as follows:

A CMC element $cmc$ may have a set of input failure modes $IFM = \left\{ ifm_1,...,ifm_q \right\}$ which represent incoming failures from outside of the component's scope with the rate $\tau(ifm)$ where $ifm \in IFM$. Each input failure mode $ifm \in IFM$ can affect one or several transitions $T_{ifm} \subseteq T$ of the CMC.
This relation is represented by a set of so-called \emph{Input Failure Mode Dependencies}:
\begin{equation}
DI = \left\{ (ifm, t) \; | \; ifm \in IFM, t \in T \right\}
\end{equation}
The dependency of one transition $t \in T$ is defined as 
\begin{equation}
DI(t) := \left\{ ifm \; | \;  (ifm,t) \in DI \right\}
\end{equation}
When a transition $t(a,b) \in T$ from state $a \in S$ to $b \in S$ is interconnected with one or more input failure modes, the rate of the transition $t(a,b)$ changes to
\begin{equation}
t(a,b) := \tau_{a,b} + \sum_{ifm \in DI(t)} \tau_{ifm}
\end{equation}
                                 
Moreover, a CMC may have a set of output failure modes $OFM = \left\{ ofm_1,...,ofm_r \right\}$ which represent failures propagated to other components with the failure rate $\tau(ofm)$ of $ofm \in OFM$. 

The rate $\tau(ofm)$ is calculated as the rate that a corresponding specific error state $s \in S_{error}$ is reached, since the error states of the CMC represent the failures modes.
This relation is represented by a set of so-called \emph{Output Failure Mode Dependencies}:
\begin{equation}
DO = \left\{ (s, ofm) \; | \; s \in S, ofm \in OFM \right\}
\end{equation}
If the error state $s$ is reached, the output failure mode $ofm$ is triggered. Thus, $\tau(ofm)$ is the calculated rate $\tau(s)$ for arriving in error state $s$.

Hence, a Component Markov Chain (CMC) can be defined by the tuple
\begin{equation}
CMC = \left( S, S_{error}, s_{init}, T, IFM, DI, OFM, DO\right)
\end{equation}

\begin{figure}[htbp]
  \centering
  \includegraphics[width=0.7\linewidth]{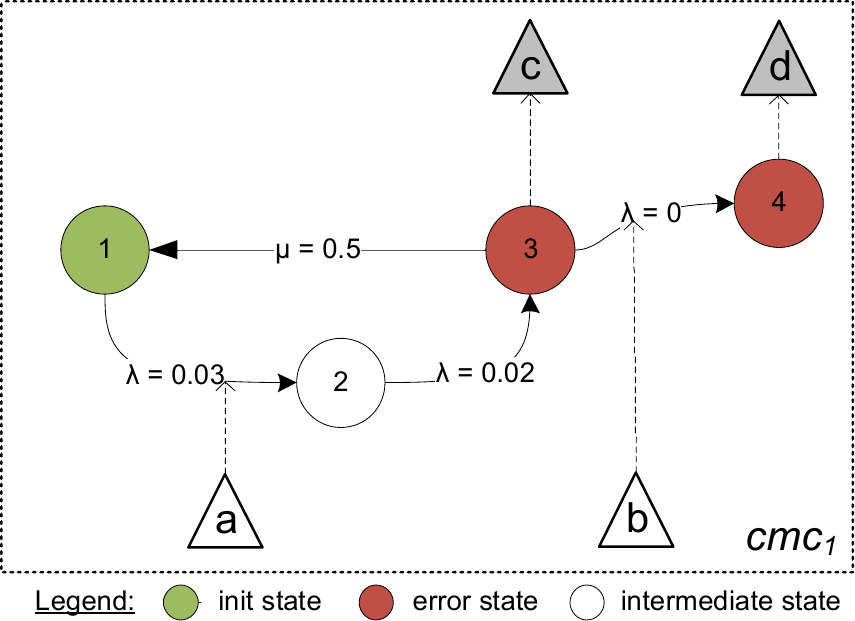}
  \caption{Exemplary Component Markov Chain (CMC)}
  \label{fig:example_CMC}
\end{figure}

An example for a CMC is illustrated in Fig.~\ref{fig:example_CMC}.
This exemplary MC is defined by:
\begin{IEEEeqnarray*}{rCl}
S &=& \left\{ 1,2,3,4 \right\} \\
s_{init} &=& 1 \\
S_{error} &=& \left\{ 3,4 \right\} \\
T &=& \left\{ (1,2),(2,3),(3,1),(3,4) \right\} \\
IFM &=& \left\{ a,b \right\} \\
DI &=& \left\{ (a,t_1), (b,t_4) \right\} \text{with } t_1:=(1,2), t_4:=(3,4) \\
OFM &=& \left\{ c,d \right\} \\
DO &=& \left\{ (3,c), (4,d) \right\}
\end{IEEEeqnarray*}

Hence, the rates of the transitions of the exemplary CMC are as follows:
\begin{IEEEeqnarray*}{rCl}
t(1,2) := \tau_{1,2} &=& \lambda_{1,2} + \tau(a) = 0.03 + \tau(a) \\
t(2,3) := \tau_{2,3} &=& \lambda_{2,3} = 0.02 \\
t(3,1) := \tau_{3,1} &=& \mu_{3,1} = 0.5 \\
t(3,4) := \tau_{3,4} &=& \lambda_{3,4} + \tau(b) = 0.0 + \tau(b) = \tau(b)
\end{IEEEeqnarray*}

When the CMC is analyzed, the values for $\tau(c)$ and $\tau(d)$ are calculated.

A CMC $cmc$ can be associated to a system component $c \in C$ in the same way a CFT element is associated with a component \cite{Domis2008}:
$\tilde{CMC}(c) = cmc$. 

Thereby, it is possible to map the input \& output failure modes of a CMC $cmc$ to the input \& output ports of the component $c$ via the function $map$.
Hence, each input failure mode $ifm \in IFM$ is related to an inport $in \in IN(c)$ of the component $c \in C$:
\begin{equation}
  \begin{split}
		\forall \; ifm& \in IFM: \\
									  & \exists \; in \in IN(c) \text{ with } map(ifm) = in
	\end{split}
\end{equation}
and each output failure modes $ofm \in OFM$ is related each to an outport $out \in OUT(c)$ of the component $c \in C$:
\begin{equation}
  \begin{split}
		\forall \; ofm& \in OFM: \\
										& \exists \; out \in OUT(c) \text{ with } map(ofm) = out
	\end{split}
\end{equation}

Based on the construction of the CMC it is possible to combine CMC and CFT elements and analyze it qualitatively and quantitatively.

\section{Generalized Hybrid Component Fault Trees}
\label{sec:hyCFT}
A first concept to combine CFTs and MCs was already introduced in \cite{4224161}. However, the use of MCs in this approach was limited to the substitution of basic events with one failure state. Hence, MCs can only be used to model the failure behavior of a sensor.
With the generalized component concept for MCs as presented in Sec.~\ref{sec:cmc}, GHCFTs can be built from any combination of CFT and CMC elements. 
The only limitation in building a GHCFT is that common cause failures (repeated events within Fault Trees) can only be input to one CMC, if this captures the whole influence of this repeated event on the top event of the GHCFT.

\begin{figure}[htbp]
  \centering
  \includegraphics[width=1.0\linewidth]{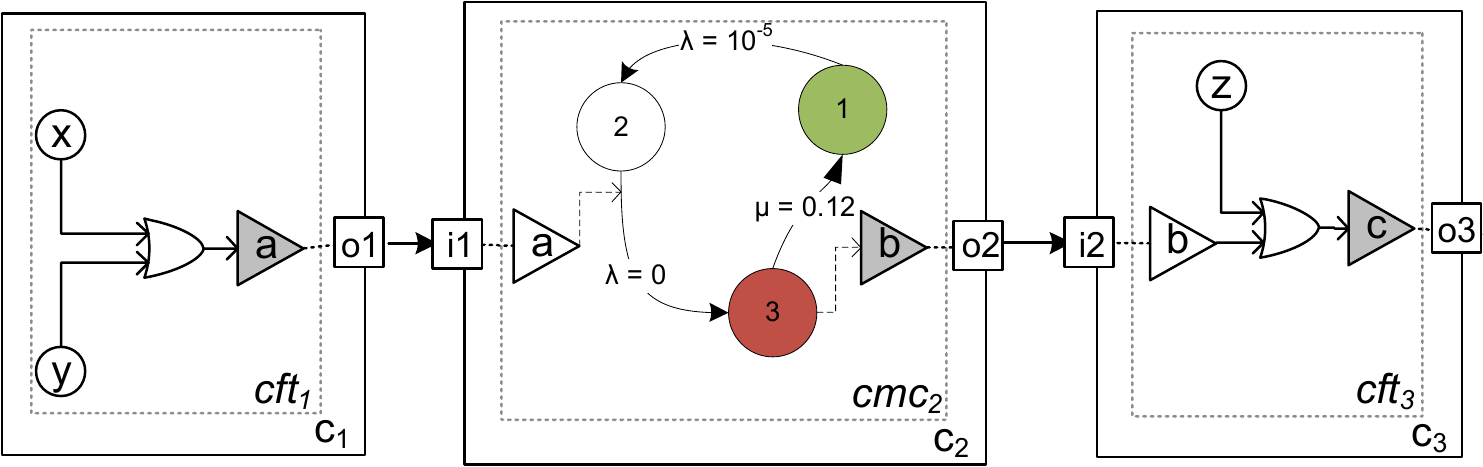}
  \caption{Exemplary Generalized Hybrid Component Fault Tree (GHCFT)}
  \label{fig:example_hft}
\end{figure}

The $GHCFT(S)$ of a system $S$ is defined a follows:
Each component $c \in C$ has a failure logic model $flm$, which is either a CFT element $cft_c \in CFT$ or a CMC element $cmc_c \in CMC$ representing the component's failure behavior:
\begin{equation}
\forall \; c \in C: flm(c) = \left\{ 
\begin{array}{l l}
  cft_c \in CFT \\
  cmc_c \in CMC \\ \end{array} \right.
\end{equation}
Fig.~\ref{fig:example_hft} presents an exemplary GHCFT.

This example is defined by:
\begin{IEEEeqnarray*}{rCl}
C &=& \left\{ c_1,c_2,c_3 \right\} \\
IN(c_1) &=& \left\{ \right\} \\
IN(c_2) &=& \left\{ i1 \right\} \\
IN(c_3) &=& \left\{ i2 \right\} \\
OUT(c_1) &=& \left\{ o1 \right\} \\
OUT(c_2) &=& \left\{ o2 \right\} \\
OUT(c_3) &=& \left\{ o3 \right\} \\
CON &=& \left\{ (o1,i1),(o2,i2) \right\} \\
flm(c_1) &=& cft_1 \\
flm(c_2) &=& cmc_2 \\
flm(c_3) &=& cft_3
\end{IEEEeqnarray*}

With GHCFTs there is a modular and hierarchical approach to describe the failure behavior of a system. The concept of GHCFT is more powerful in terms of expressiveness than classical Fault Tree or CFT methodology, because it is possible to express dynamic system behavior. Since a GHCFT can be composed of CFT and CMC elements, the failure behavior of complex, large-scale systems can be modeled using a divide-and-conquer strategy. Hence, state explosion is avoided, since not the complete system has to be modeled using a MC, but only the part with relevant dynamic or state-dependent behavior (e.g.~fault-tolerant components).
Moreover, CMC elements can be composed automatically in the same way as CFT elements can be composed (cf.~\cite{Moehrle2016}).

Thus, a GHCFT has a similar power w.r.t. expressiveness as \emph{AltaRica 3.0} \cite{6622976} or \emph{State/Event Fault Trees (SEFTs)} \cite{Kaiser2007} regarding the attributes identified in \cite{Lipaczewski2015191} (see table \ref{tab:comparison}). The GHCFT methodology enables the modeling of the failure behavior of a system in dynamic, event-based way by using states \& transition in CMC elements and it allows the user to specify and compose hierarchical systems by using the principles of the CFT methodology.  
Moreover, a GHCFT can be analyzed with the well-known algorithms for qualitative and quantitative FTA \cite{RUIJTERS201529}. Thereby, our approach avoids a potential explosion of state-spaces and thus performance problems during analysis, which is according to \cite{sharvia2015model} the main drawback of the AltaRica as well as the SEFT methodology. 

\begin{table}[htbp]
\caption{Comparison GHCFT, AltaRica 3.0 and SEFT}
\begin{tabular}{|l|c|c|c|}
\hline
\textbf{Comparison} & \textbf{GHCFT} & \textbf{AltaRica} & \textbf{SEFT}\\
\textbf{criteria} 	& 							 &  								 & \\
\hline
Event based 				& Yes 					 & Yes 							 & Yes\\
\hline
Composition 				& Yes 					 & Yes 							 & Yes\\
\hline
Hierarchy 					& Yes 					 & Yes 							 & Yes\\
\hline
Graphical 					& Yes 					 & Yes							 & Yes\\
representation 			& 							 & 									 & \\
\hline
Analysis  					& qualitative 	 & Compilation to 	 & Transformation \\
methods	  					& and		  	  	 & Fault Trees and 	 & to Deterministic \\
										& quantitative	 & Markov graphs,		 & and Stochastic \\
										&	FTA				  	 & stochastic and  	 & Petri Nets \\
										&								 & stepwise 				 & \\
										&								 & simulation				 & \\
\hline
\end{tabular}
\label{tab:comparison}
\end{table}


\subsection{Qualitative Analysis}
\label{subsec:mcs}
For qualitative analysis the CMC is transformed into a CFT element. This transformation is performed in two steps:
\begin{enumerate}
\item	For each transition between two states a basic event is created. If the transition has an input failure mode dependency then an OR gate is created and all IFMs the transition is depending on as well as the basic event of the transition are connected to this OR gate. 
\item For each OFM of the CMC and for each path leading from the initial state of the CMC to an error state connected to an OFM, the Basic Events or OR gates with connected IFMs within the CFT element, which represent the transition, are connected by an AND gate. If more than one path is existing, all AND gates are connected by an OR gate, which is then connected to the OFM.
\end{enumerate}

\begin{figure}[htbp]
  \centering
  \includegraphics[width=0.75\linewidth]{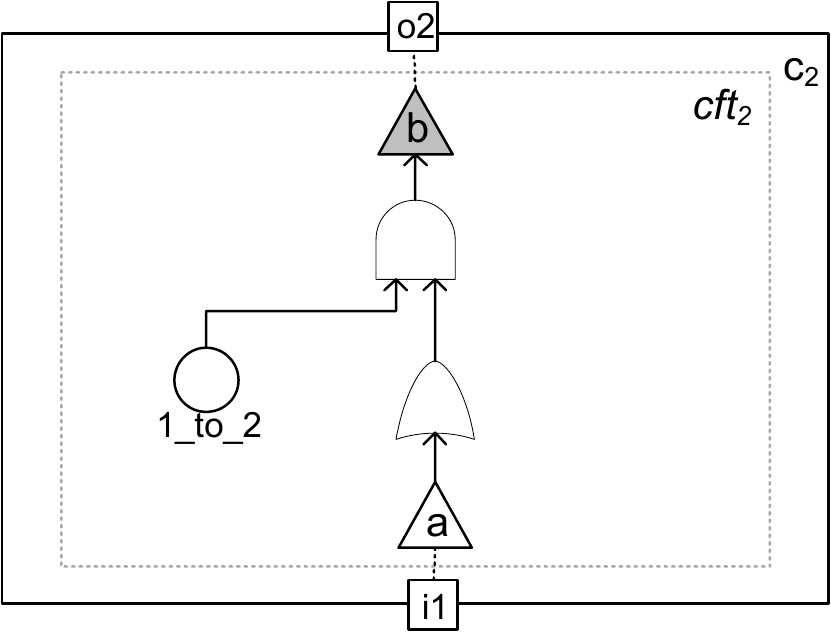}
  \caption{Exemplary CMC transformed into a CFT element for qualitative analysis}
  \label{fig:example_hFT_qualitative}
\end{figure}

For the exemplary GHCFT as depicted in Fig.~\ref{fig:example_hft}, the CMC $cmc_2$ for the qualitative analysis is transformed into a CFT as presented in Fig.~\ref{fig:example_hFT_qualitative}.


\subsection{Quantitative Analysis}
The failure rate $\tau(ofm)$ is calculated in the Markov Chain-Analysis: A general analysis approach is to use an implicit numerical integration scheme with step-size control for the stiff system of ordinary differential equations representing the CTMC as described e.g. in \cite{476004}. This also allows non-constant transition rates at inports resulting from redundancy in upstream CFTs or CMCs.
Thus, the failure rate of each OFM of a CMC can be calculated and the CMC can be combined with CFTs.

Since the failure rate of each OFM of a CMC can be calculated as described above, CMC can be combined with CFT elements and integrated into a \emph{GHCFT}. Please note, that common cause failures represented as repeated basic events within the GHCFT can be input to one CMC, if this captures the whole influence of this repeated event on the top event of the GHCFT. Otherwise the GHCFT cannot be analyzed quantitatively.

For the exemplary GHCFT as depicted in Fig.~\ref{fig:example_hft}, assuming an infinite mission time, we need no numerical integration and can, however, explicitly calculate $\tau(c)$ the stationary solution in the following way.
Thereby, we assume the following failure rates for the basic events in Fig.~\ref{fig:example_hft}:
\begin{IEEEeqnarray*}{rCl}
\lambda_x &=& 2.0 * 10^{-7} \\ 
\lambda_y &=& 4.0 * 10^{-7} \\ 
\lambda_z &=& 1.0 * 10^{-7}
\end{IEEEeqnarray*}
Thus, the failure rate of the input failure mode $a$ is $\tau(a) = 6.0 * 10^{-7}$, and the quantitative analysis of the CMC $cmc_2$ can use the reciprocal of the sum of mean durations of staying in states 1 and 2 to calculate:
\begin{IEEEeqnarray*}{rCl}
\tau(b) = \tau(3) &=& \frac{1}{\frac{1}{\tau(1,2)} + \frac{1}{\tau(2,3)}} = \frac{1}{\frac{1}{1.0 * 10^{-5}} + \frac{1}{0 + \tau(a)}} \\
            &=& \frac{1}{\frac{1}{1.0 * 10^{-5}} + \frac{1}{6.0 * 10^{-7}}} \\
						&=& 5.66 * 10^{-7}
\end{IEEEeqnarray*}

Based on this result, the failure rate of the top event $c$ of the GHCFT is calculated as follows:
\begin{IEEEeqnarray*}{rCl}
\tau(c) &=& \lambda_z + \tau(b) \\
		    &=& 1.0 * 10^{-7} + 5.66 * 10^{-7} = 6.66 * 10^{-7}
\end{IEEEeqnarray*}
Please note, that with this approach repeated events can only be handled as input of one CMC, if this input captures the whole influence of this repeated event on the top event of the GHCFT.

\section{Case Study}
\label{sec:casestudy}
In this section, a case study from the automotive domain is presented to demonstrate our approach. 
The presented system is a radio controlled car. 
Please note, that this system is a simplification of a real system to be used for demonstration purposes.
%

\begin{figure}[htbp]
  \centering
	\includegraphics[width=1.0\linewidth]{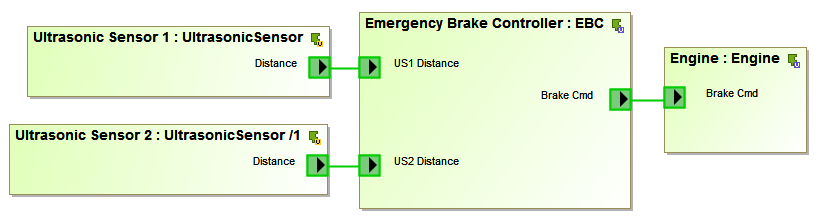}
  \caption{System architecture as a SysML Internal Block Diagram}
  \label{fig:casestudy_architecture}
\end{figure}

\begin{figure}[htbp]
  \centering
  \includegraphics[width=1.0\linewidth]{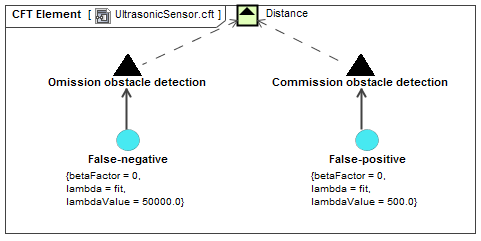}
  \caption{CFT element of an Ultrasonic Sensor component}
  \label{fig:casestudy_cft}
\end{figure}

\begin{figure*}[htbp]
  \centering
  \includegraphics[width=1.0\linewidth]{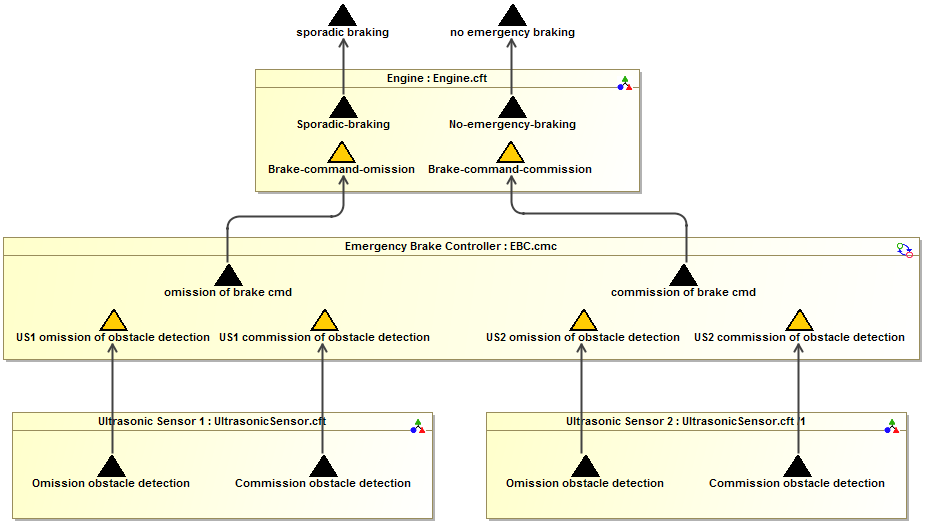}
  \caption{Generalized Hybrid Component Fault Tree (GHCFT) of the Emergency Braking Function}
  \label{fig:casestudy_hycft}
\end{figure*}

Fig.~\ref{fig:casestudy_architecture} shows the system architecture of an emergency braking functionality in the radio controlled car in form of a \emph{SysML Internal Block Diagram (IBD)}.
The architecture consists of the components \emph{Ultrasonic Sensor 1 (US 1)}, \emph{Ultrasonic Sensor 2 (US 2)}, \emph{Emergency Braking Controller (EBC)} (implemented in software) and \emph{Engine (E)}, which exchange data via the modeled ports and interconnections.
Both sensors are a source for random faults leading to the hazardous situations on system level (cf.~Fig.~\ref{fig:casestudy_cft} \& Fig.~\ref{fig:casestudy_hycft}). The random failures are represented by the basic events \emph{false-negative} with a failure rate of 50000 FIT (Failures in Time, in $10^9$h) and \emph{false-positive} with a failure rate of 500 FIT in the CFT element of the sensor component.
The omission failure mode (\emph{omission obstacle detection} / \emph{no emergency braking}) refers to the situation where accidentally no braking command is generated by the EBC component. The commission failure mode (\emph{commission obstacle detection} / \emph{sporadic braking}) refers to the situation where accidentally a braking command is generated.

\begin{figure*}[htbp]
  \centering
  \includegraphics[width=1.0\linewidth]{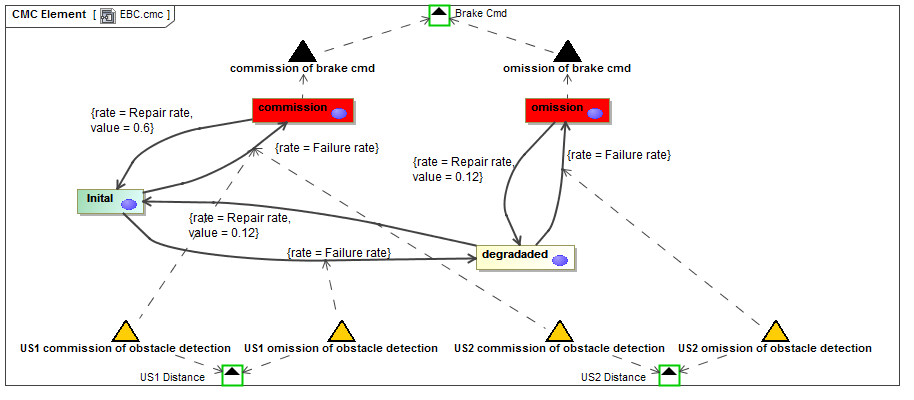}
  \caption{CMC element of the Emergency Braking Controller software component}
  \label{fig:casestudy_cmc}
\end{figure*}

The EBC component sends a brake command to the electric engine E, if one of the sensors detects an obstacle in front of the vehicle. The controller is fault-tolerant in the sense, that the system can still determine an obstacle if only US 1 is defect. US 2 is a cold spare of US 1. 
The failure rate of US 2 in stand-by mode prior to its activation after the failure of US 1, is neglected. Therefore, we look at the failure of US 2 after the failure of US 1 leading to the overall failure of the emergency braking function.
Since such temporal sequences of events behavior cannot be expressed using a CFT element, it is represented using CMC element as described in Sec.~\ref{sec:cmc} (cf.~Fig.~\ref{fig:casestudy_cmc}).
Fig.~\ref{fig:casestudy_hycft} shows the GHCFT representing the failure behavior of the emergency braking function in our case study. It consists of a CMC element for the EBC and CFT elements for the Ultrasonic Sensors as well as the Engine component.
The GHCFT can be analyzed qualitatively and quantitatively as outlined in Sec.~\ref{sec:hyCFT}.

The results of the qualitative and quantitative analysis of the GHCFT presented in Fig.~\ref{fig:casestudy_hycft} are as follows:
\begin{table}[htbp]
\caption{Results of analysis of the GHCFT defined for the case study}
\begin{tabular}{|l|l|c|c|}
\hline
\textbf{Top event} & \textbf{Cut sets} & \textbf{Failure rate} & \textbf{MTBF}\\
									 & 									 & \textbf{[fit]} & \textbf{[h]}\\
\hline
sporadic & (1) US 1.False-positive & $999.58$ & ~$1 \cdot 10^7$\\
braking	 & (2) US 2.False-positive & & \\
\hline
no emergency & (1) US 1.False-negative, & $20.82$ & $4.80 \cdot 10^7$ \\
braking	     & \quad\; US 2.False-negative & & \\
\hline
\end{tabular}
\label{tab:results}
\end{table}


\section{Conclusions}
\label{sec:summary}
In this paper, we present a concept to integrate Markov Chains in Component Fault Tree models. Based on a component concept for Markov Chains which allows the modular specification of a Markov Chain and its association to a system development element such as a component similar to Component Fault Tree (CFT) methodology. Thus, complex or software-controlled systems can be analyzed w.r.t.~safety or reliability in a modular and compositional way.
The combination of CFT and CMC models results in a so-called \emph{Generalized Hybrid Component Fault Tree (GHCFT)}, which can then be analyzed qualitatively (Minimal Cut Set analysis) or quantitatively using well-established algorithms. This combination of CMC and CFT models allows less complex quantitative analysis that takes the temporal sequence of failure events or state-dependent behavior into account compared to an analysis using a Markov Chain for the whole system. The only limitation of our approach is that in case a common cause failure enters a CMC, its effect at the CMC inport must already represent the complete effect on the top event of the GHCFT.

Future work will be a more detailed evaluation using a large-scale industrial case study which shows the benefits of our approach in analyzing safety or reliability. Moreover, we will investigate approaches how to handle repeated events within GHCFTs.

\section*{Acknowledgment}
Parts of the work leading to this paper was funded by the Framework Programs for Research and Innovation Horizon 2020 under grant agreement No.~732242 (DEIS).

\bibliographystyle{IEEEtran}
\bibliography{paper_references}

\end{document}